\documentclass[twocolumn]{aastex63}
\usepackage{commath}
\usepackage{epsfig}

\accepted{ in ApJL}

\shorttitle{Pop III stars }
\shortauthors{Latif  et al.}

\begin{document}

\title{Magnetic braking during direct collapse black hole formation}



\correspondingauthor{Muhammad A. Latif}
\email{latifne@gmail.com}

\author{Muhammad A. Latif}
\affiliation{Physics Department, College of Science, United Arab Emirates University, PO Box 15551, Al-Ain, UAE}

\author{Dominik R.G. Schleicher}
\affiliation{Astronomy Department, Universidad de Concepci\'on, Barrio Universitario, Concepci\'on, Chile}

\begin{abstract}
Magnetic fields are expected to be efficiently amplified during the formation of the first massive black holes via the small-scale dynamo and in the presence of strong accretion shocks occurring during gravitational collapse. Here, we analyze high-resolution cosmological magneto-hydrodynamical simulations of gravitational collapse in atomic cooling halos, exploring the dynamical role of magnetic fields, particularly concerning the effect of magnetic braking and angular momentum transport. We find that after the initial amplification, magnetic fields contribute  to the transport of angular momentum and reduce it compared to pure hydrodynamical simulations. However, the magnetic and Reynolds torques do not fully compensate for the inward advection of angular momentum, which still accumulates over timescales of $\sim1$~Myr. A Jeans analysis further shows that magnetic pressure strongly contributes to suppressing fragmentation on scales of $0.1-10$~pc. Overall, the presence of magnetic fields thus aids in the transport of angular momentum and favors the formation of massive objects.
\end{abstract}


\section{Introduction} \label{sec:intro}
Magnetic fields are ubiquitous throughout the cosmos and are considered responsible for various astrophysical phenomena, such as the transfer of angular momentum, launch of collimated jets and outflows, suppressing fragmentation, and stabilizing accretion disks \citep{Beck99,Beck07,Pud12}. They may have a primordial origin \citep[e.g.][]{Widrow2012} or result from the efficient amplification of weak seed fields due to astrophysical dynamos \citep[e.g.,][]{Brandenburg2005}. It has been suggested that magnetic fields  may play an important role already during the formation of the first objects in the Universe, particularly the first stars and supermassive black holes \citep[e.g.][]{Pudritz1989, Sethi2005, Silk2006, Schleicher2009}. 
The origin of magnetic fields is still uncertain as the standard model does not provide any constraints on their strength. Magnetic field may have generated during the cosmic inflation through electro-weak or quantum chromodynamics phase transitions or alternatively via the Biermann battery effect and the Weibel instability (see review by \cite{Widrow2012}). The current observational constraints on the strength of intergalactic magnetic fields are derived from CMB observations which were suggested to provide an upper limit of a few nano Gauss while blazer observations provide lower limit of $\rm 10^{-16}$ G \citep{Kahn10, Nero10,P16}.  Irrespective of their origin, the seed magnetic fields are many orders of magnitude smaller than the present day fields.

Over the last decade, numerous studies have suggested that magnetic fields, irrespective of their initial field strength, can be efficiently amplified by the small scale dynamo during first structure formation \citep{Schleicher10,Sur10,Souza2010,Fed11,Schobera,Turk2012,Latif13m,Grete19}. Furthermore, they can get amplified via the $\alpha-\Omega$ dynamo in the presence of rotation \citep{ls15, Sharda20} and strong accretion shocks  due to the rapid infall within cosmological simulations \citep{Latif14mag,Hir21,Hir22}. Such strongly amplified fields inhibit fragmentation  and stabilize the central accretion disks \citep{Latif23} (hereafter L23), as also seen in \cite{Hir21}. L23 performed  cosmological magneto-hydrodynamical (MHD) simulations in the context of direct collapse black hole formation, evolving them for about 1.6 Myr, a timescale comparable to the lifetime of supermassive stars \citep{Janka2002}. This study focused  on assessing the impact of magnetic fields on the degree of fragmentation and the masses of clumps by comparing their results with hydrodynamical runs.

In the context  of the first massive objects, the potential role of magnetic fields in  suppressing fragmentation via the magnetic Jeans mass has been suggested  previously \citep[e.g.][]{Schleicher2009, Latif2016}, along with the presence of the magneto-rotational instability to drive accretion within the disk \citep{Silk2006}. However, as it is well-known in the context of present-day star formation, magnetic fields can also affect angular momentum transport and help to delay or suppress the formation of rotationally supported disks \citep{Kulsrud1971, Galli2006, Shu2007, Hennebelle2016, Sheikhnezami2022}. It is  not clear how  magnetic torques compare to the Reynold torques and what are their relevant contributions. In this letter, simulations by L23 are analyzed with respect to the magnetic braking and its effect on  distribution of the angular momentum within the collapsing halos. A short summary of the methods employed to perform the previous simulations is given in section~\ref{sec:methods}. The results of the analysis are presented in section~\ref{sec:results} and a final discussion and conclusions are provided in section~\ref{sec:conc}.

\begin{figure*} 
\begin{center}
\includegraphics[scale=0.5]{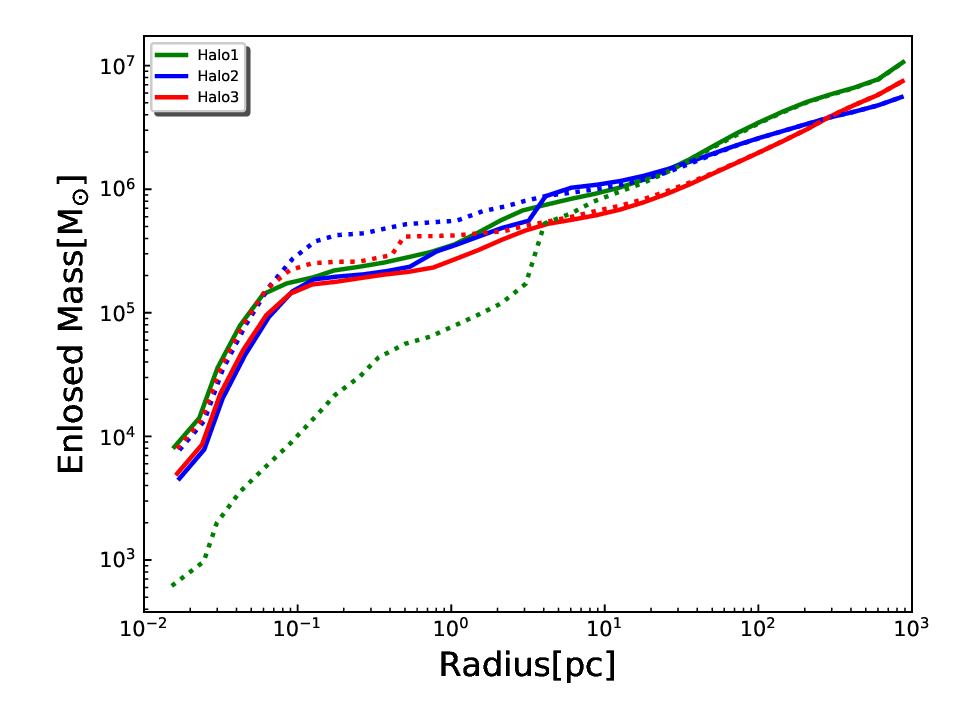}
\includegraphics[scale=0.5]{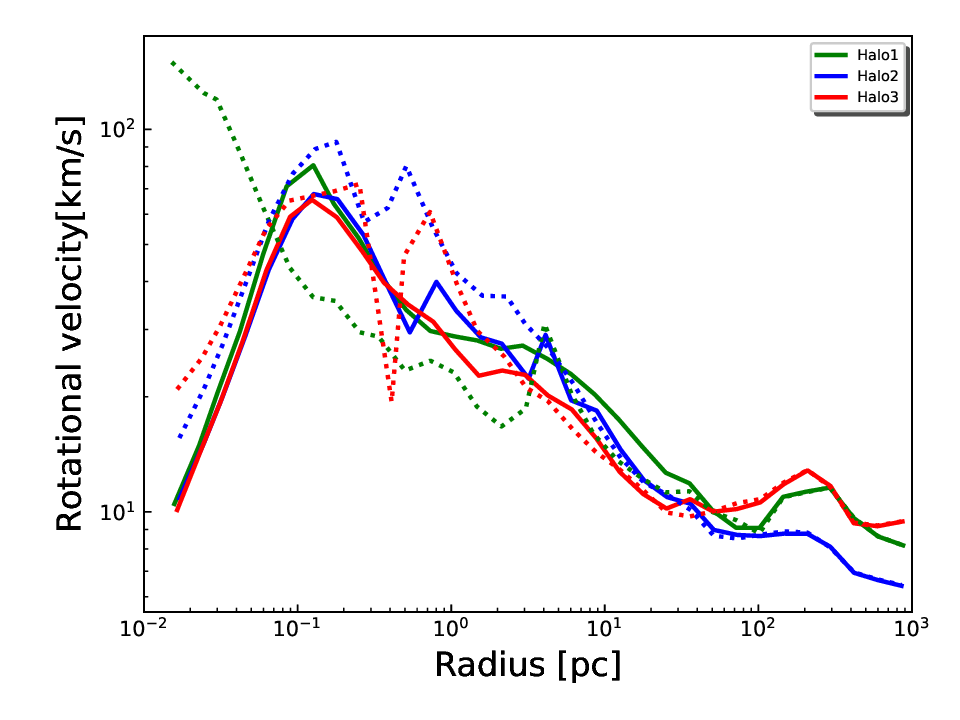}
\includegraphics[scale=0.5]{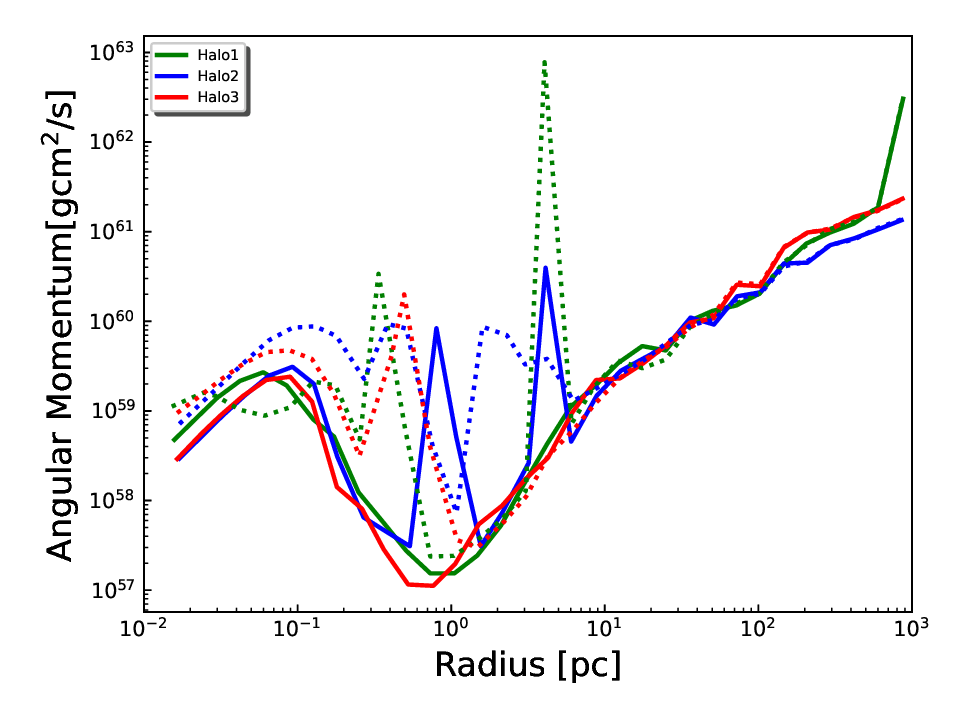} 
\includegraphics[scale=0.5]{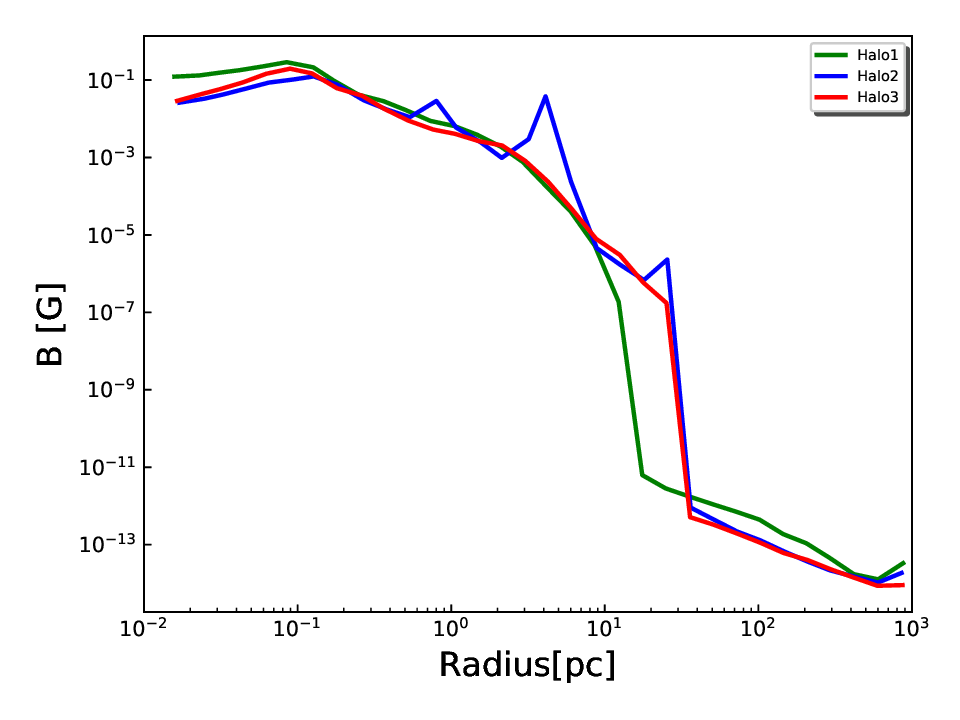} 
\includegraphics[scale=0.5]{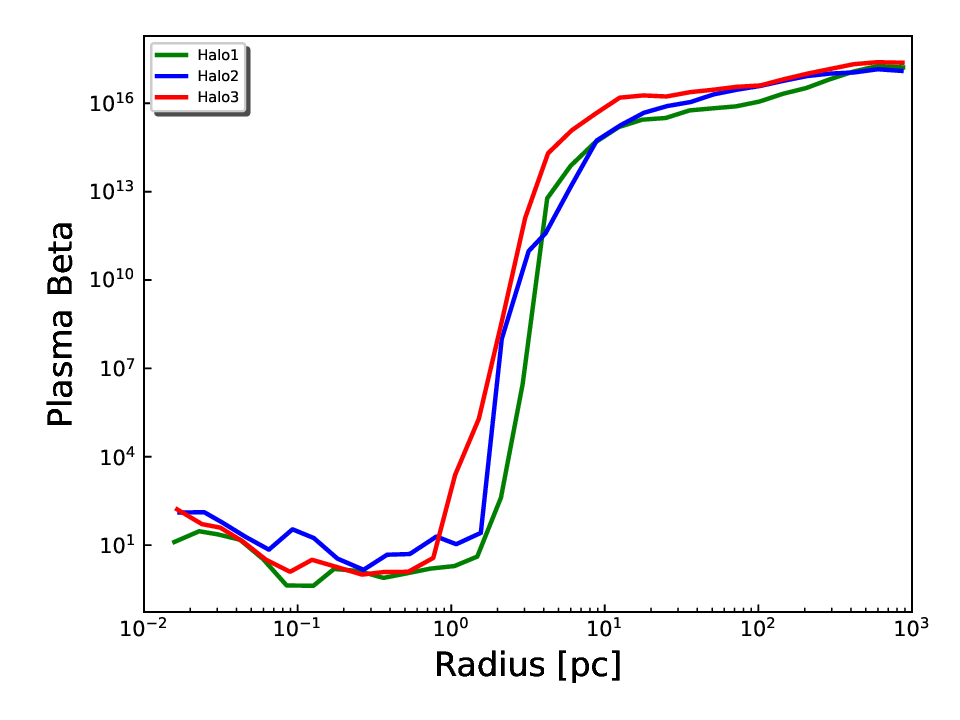} 
\includegraphics[scale=0.5]{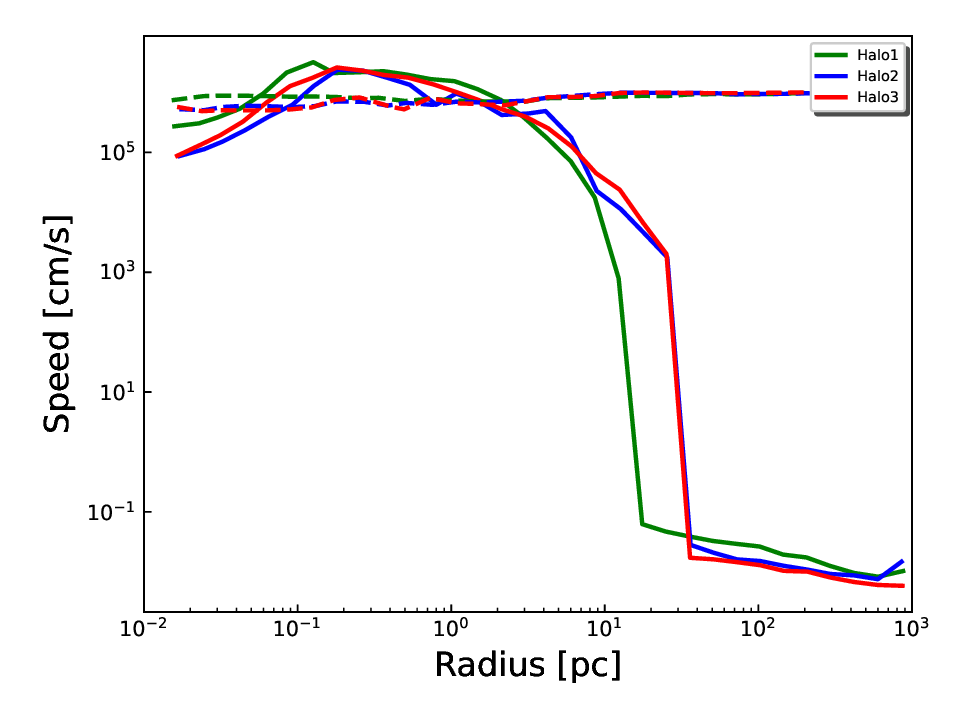} 
\end{center}
\vspace{-0.1cm}
\caption{Mass-weighted radial profiles of the three simulated halos. Top left: Enclosed mass. Top right: Rotational velocity. Mid left: Angular momentum. Mid right: Magnetic field strength (in proper units). Bottom left: Plasma beta parameter. Bottom right: Alfv\'en and thermal velocity. The solid lines correspond to the MHD simulations, the dotted lines to the hydrodynamical ones. In the bottom right figure the the dashed line refers to the sound speed, which we checked to be very similar in the MHD and hydrodynamical simulations.}
\label{fig:f1}
\end{figure*}

\begin{figure} 
\begin{center}
\includegraphics[scale=0.5]{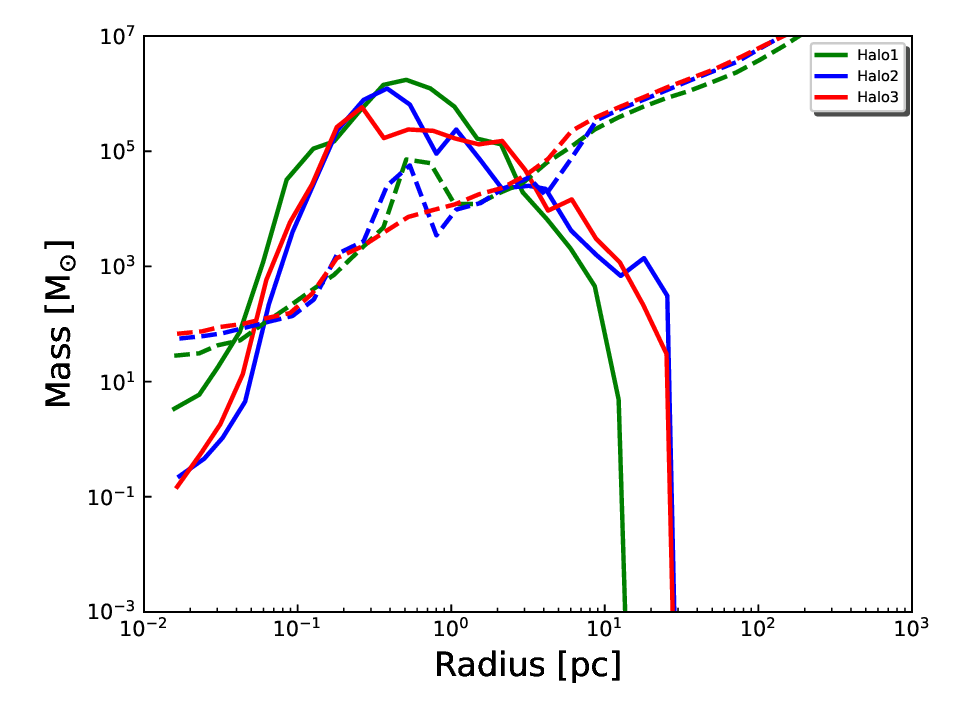}
\end{center}
\vspace{-0.1cm}
\caption{Mass-weighted radial profiles of the thermal (dashed line) and magnetic Jeans mass (solid line) for the three simulated halos.}
\label{fig:Jeans}
\end{figure}

\begin{figure*} 
\begin{center}
\includegraphics[scale=1.0]{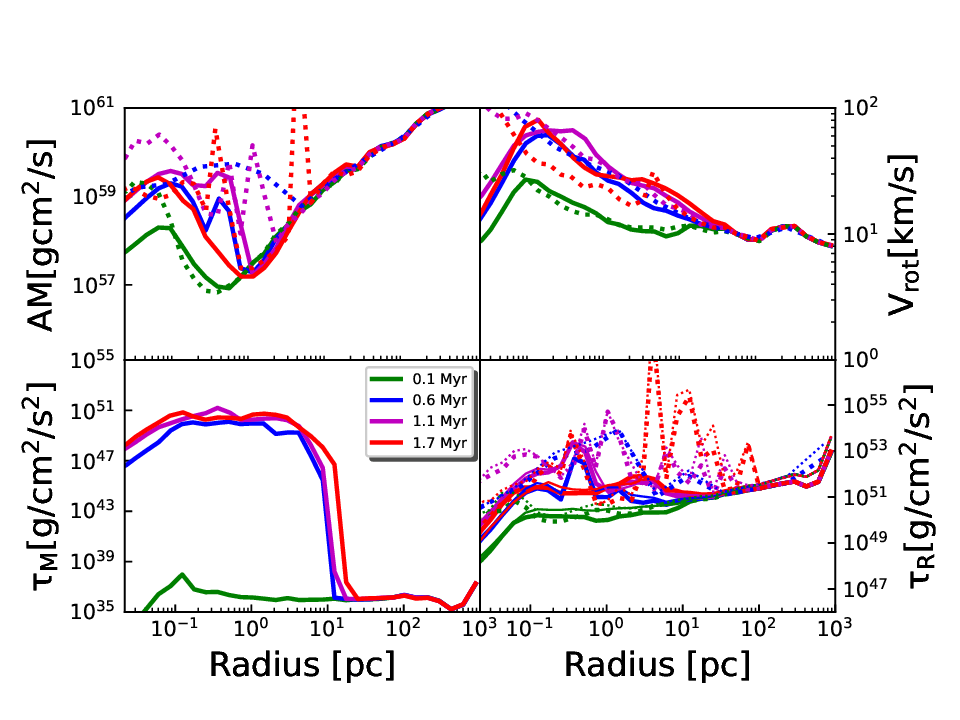} 
\end{center}
\vspace{-0.1cm}
\caption{Mass-weighted radial profiles of halo 1. Top: Time evolution of angular momentum and rotational velocity in the MHD (solid line) and hydrodynamical simulations (dotted line) of halo 1. Bottom: Time evolution of Reynolds and magnetic torque. The Reynolds torque is shown for the MHD (solid line) and hydrodynamical (dotted line) simulations. We also provide a comparison with the inward advection term given in Eq.~\ref{Jadv} (thin lines both dotted and solid).
}
\label{fig:ang}
\end{figure*}

\begin{figure*} 
\begin{center}
\includegraphics[scale=1.0]{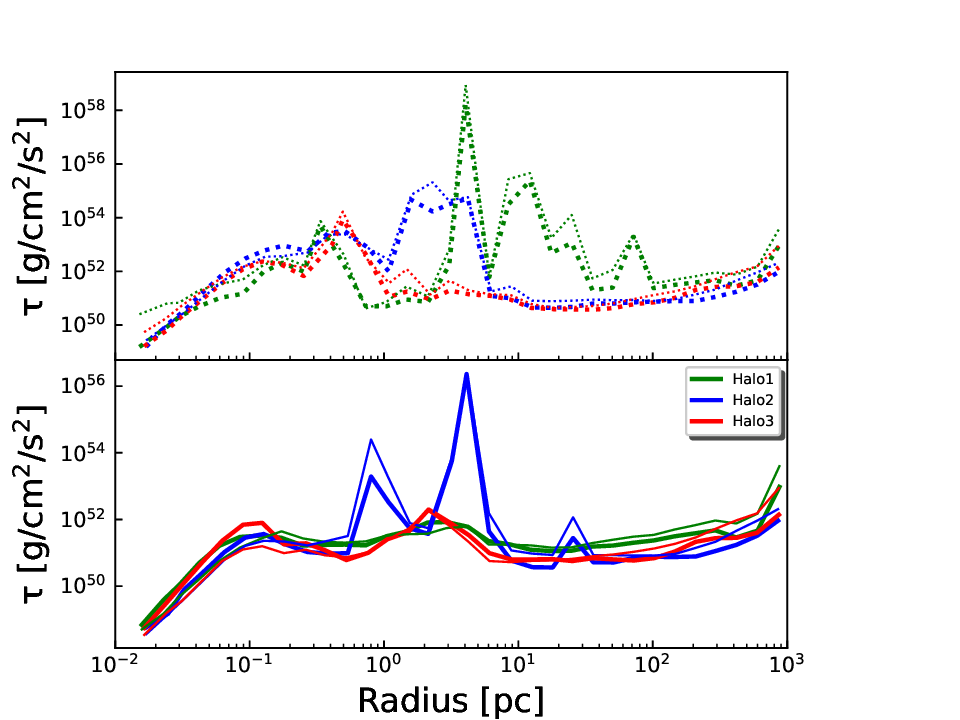} 
\end{center}
\vspace{-0.1cm}
\caption{The top panel is  showing the mass-weighted radially averaged Reynold torques for HD runs and the bottom panel the total torque (Reynolds plus magnetic) for the MHD runs at the end of simulations for all three halos. For comparison, the thin lines in both panels show the angular momentum inflow term due to advection, $\dot{J}_{\rm adv}$ in both panels. }
\label{fig:torque}
\end{figure*}

\begin{figure} 
\begin{center}
\includegraphics[scale=0.54]{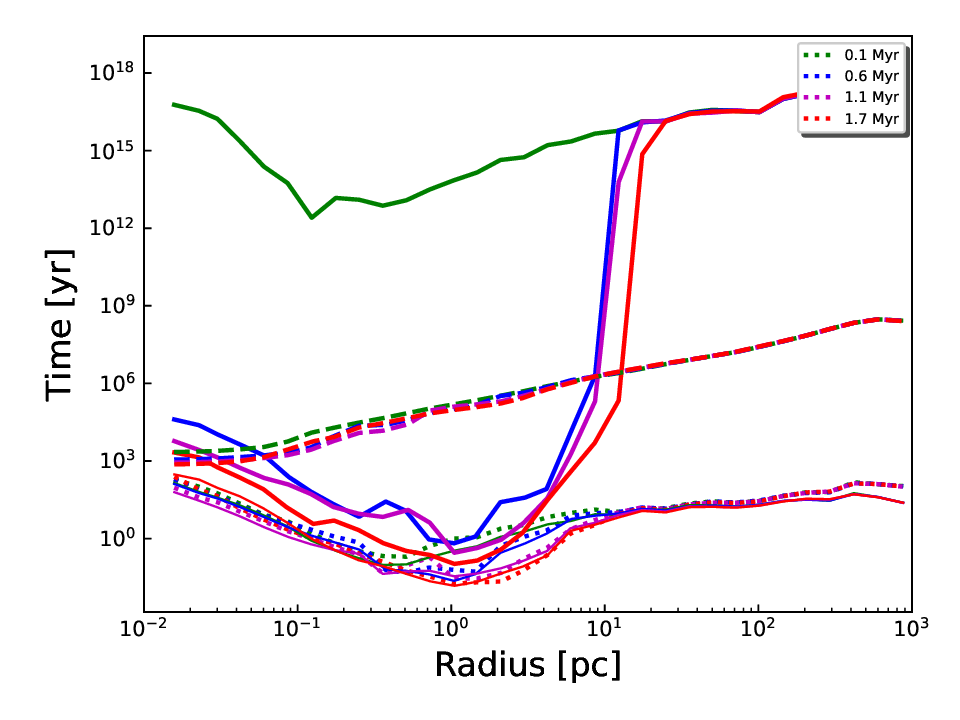}
\end{center}
\vspace{-0.1cm}
\caption{Mass-weighted radially averaged profile of the characteristic timescales associated with Reynolds torque (dotted lines), magnetic torque (solid lines), infall torques (thin solid lines) in comparison to the free-fall (dashed line) timescale. }
\label{fig:torque3}
\end{figure}

\section{Numerical Method} \label{sec:methods}
We analyzed here the cosmological magnetohydrodynamics simulations   performed with adaptive mesh refinement code \textsc{ENZO} published in L23. A brief summary of these simulations is presented here and for further details interested readers are referred to L23. Cosmological magnetohydrodynamical simulations were performed for three distinct halos and their results were compared with hydrodynamical runs. The simulated halos had a masses of $\rm 3 \times 10^7~M_{\odot}$, $\rm 1.7 \times 10^7~M_{\odot}$ and $\rm 2.3 \times 10^7~M_{\odot}$ at z= 13.2, 12.5 \& 12.6, with respective spin parameters of 0.07, 0.01 \& 0.02. They were seeded with an initial uniform magnetic field strength of $\rm 10^{-14}~G$ ($\rm4.5 \times 10^{-19}$ G in comoving units) at z=150. The motivation for the choice of such initial field strength comes from theoretical works which predict B fields of strength $ \rm 10^{-17} - 10^{-20} $ G  at galactic scales during electroweak phase transitions \citep{Baym96,Gras01} and $\rm10^{-20}$ G from the quantum chromodynamics in the very early universe \citep{Sigl97}. We further assume uniform B field for the sake of a simplicity and coherent fields on larger scales may be generated by the $\rm \alpha-\Omega$ dynamo in the presence of helicity. All simulations have an effective dark matter resolution of $\rm \sim 67~ M_{\odot}$ and a spatial resolution of $\sim$ 2000 AU. We further ensured a minimum resolution of 64 cells per Jeans length during adaptive mesh refinement to resolve turbulent eddies \citep{Fed11,Latif13c}, using 15 (additional) refinement levels. Such a resolution allows to resolve small scale dynamo action, converting turbulent into magnetic energy \citep{Latif13m,Latif14mag}. As also shown in previous runs \citep{Latif14mag, Hir21,Hir22}, additional amplification occurs in the center of the halo when the accretion flow hits the central pressure-supported core, leading to the formation of strong shocks. The central part of the halo then becomes magnetized very efficiently.

The simulations employed a non-equilibrium chemistry network consisting of six primordial species (H, H$^+$, He, He$^+$, He$^{2+}$, and e$^-$)  which self-consistently solves the rate equations along with the (magneto-)hydrodynamics. L23 studied an isothermal gas collapse assuming that the intense Lyman Werner flux quenches the formation of $\rm H_2$ emitted by nearby star forming galaxies. Their chemical model further included cooling from collisional ionization and excitation  of H and He, radiative recombination, inverse Compton scattering and bremsstrahlung radiation. The simulations were evolved for about 1.6 Myr beyond their initial collapse employing a pressure floor technique to study the impact of magnetic fields during the formation of supermassive stars.

\section{Results} \label{sec:results}
Some of the main properties of the simulations for the three different halos are summarized in Fig.~\ref{fig:f1} via mass-weighted radial profiles. The radial profile of the enclosed mass approximately follows the form expected from an isothermal sphere on scales above $1$~pc, with the exception of occasional minor bumps due to small inhomogeneities in the density distribution. Within the central region, we first note some flattening of the enclosed mass until a radius of about $0.1$~pc, and subsequently a steep decline due to the density being approximately constant within the central region. 

On scales above $10$~pc, the rotational velocity shows the same behaviour for MHD and hydrodynamical simulations. On smaller scales it starts to deviate, with the rotational velocity being larger in two of the hydrodynamical simulations compared to the MHD ones, though there is also one simulation where this appears to be the other way round. Towards the center, the rotational velocity then usually declines, again with the exception of one halo where the center is not well-defined and other clumps are present in the vicinity, leading to more complex velocity structures. For the angular momentum, the radial profiles are very similar down to a scale of $10$~pc. On smaller scales, the hydrodynamical runs show a higher angular momentum, including strong peaks in the profile corresponding to clumps with significant amounts of angular momentum. 

In the radial profile of the magnetic field, we find signatures from compression on scales of $30-300$~pc and a steep increase around scales of $10-30$~pc, as in the inner region the magnetic field has been efficiently amplified due to turbulence and shocks as a result of strong infall. The plasma beta parameter corresponding to the ratio of thermal over magnetic pressure is initially high and of the order $10^{10}$ on scales above $10$~pc, then dropping significantly in the range of $1-10$~pc where the magnetic field strength increases very significantly, leading to typical values of $3-30$ within the central region. The sound speed in all three halos is a few times $10^5$~cm~s$^{-1}$ independent of scale, while the Alfv\'en velocity initially is insignificant of the order $0.1$~cm~s$^{-1}$ on scales above $30$~pc, though then increasing steeply and reaching values comparable to the sound speed on scales below $10$~pc. For radii of $0.1-1$~pc, it even exceeds the sound speed by a factor of $2-3$. 

This behaviour is reflected in the thermal and magnetic Jeans mass given in Fig.~\ref{fig:Jeans}. As the gas is approximately isothermal within the simulation, the thermal Jeans mass follows an approximate power-law behaviour with values of $\sim10^7$~M$_\odot$ on scales of $100$~pc  and decreasing to about $30$~M$_\odot$ on scales of $0.02$~pc, with a moderate bump on scales of $0.5$~pc where the temperature is slightly increased, as also reflected by bumps in the density structure. The magnetic Jeans mass is insignificant on scales above $30$~pc, but then rises steeply and reaches a maximum of $\sim10^6$~M$_\odot$ on scales of $\sim0.3$~pc. It dominates over the thermal Jeans mass in the range from $0.1-10$~pc and thus considerably contributes to suppress fragmentation. However, in the innermost part of the central core (scales below $0.4$~pc), the thermal support is more relevant than the magnetic one.

In Fig.~\ref{fig:ang}, we  show the time evolution of angular momentum and rotational velocity profiles of halo 1, together with magnetic and Reynolds torques given as \citep{Sheikhnezami2022}
\begin{eqnarray}
\tau_{\rm Reyn}&=&  \int_{S} r \left(\rho u_{phi}\vec{u_{p}}\right)\cdot \vec{ds},\\
\tau_{\rm M}&=& - \int_{S} r\frac{1}{4 \pi} B_{phi}\vec{B_{p}}   \cdot \vec{ds}.
\end{eqnarray}
The angular momentum is generally found to be higher in the hydrodynamical runs compared to the MHD runs. In all simulations, the angular momentum is found to increase in the center as a function of time, but more strongly within the hydrodynamical simulations. {A similar trend is found for the rotational velocity which generally increases with time. The Reynolds torque shows scatter but no significant dependence on  spatial scale and moderate increase over time, typically being in the range of $10^{51}-10^{53}$~g~cm$^{-2}$~s$^{-2}$. The magnetic torque is originally negligible as initially the magnetic field is weak, but reaches similar values of order $10^{51}$~g~cm$^{-2}$~s$^{-2}$ during the time evolution on scales less than $10$~pc. Both the Reynolds and the magnetic torques show occasional peaks in the range of $0.5-30$~pc due to inhomogeneities in the flow. We checked that halos 2 and 3 show very similar results.

We estimate the inward transport of angular momentum due to advection as\begin{equation}
\dot{J}_{\rm adv}(r)=4\pi r^2\rho r v_{\rm rot} v_r,\label{Jadv}
\end{equation}
with $\rho$ being density, $v_r$ the radial and $v_{\rm rot}$ the rotational velocity. In Fig.~\ref{fig:torque}, the sum of the magnetic and Reynolds torque is shown and compared to the advection term. Their contributions are generally found to be very similar, though the inward transport term exceeds the magnetic and Reynolds stresses at least on some scales and thus explains the inward transport of the angular momentum. Within the innermost core on scales below $0.1$~pc, though, we note that the inward transport term decreases more strongly as the innermost core is still gravitationally stable, implying lower radial velocities. Similarly, we note that the Reynolds and magnetic torques decrease in the central core due to lower magnetic fields strengths and reduced velocities on these scales.

In Fig.~\ref{fig:torque3}, we compare the timescales associated with the Reynolds and magnetic torques as well as the inward advection timescale with the  free-fall timescale, given as
\begin{equation}
T_{\rm ff}= \sqrt{\frac{3\pi}{16G\rho}} 
\end{equation}

The free-fall time follows an approximate power-law behaviour starting around $10^7$~years on scales of $1000$~pc and reaching about $10^3$~years around $0.02$~pc. The timescale related to the Reynolds stress is always considerably shorter than the free-fall time and only becomes comparable in the central region, where the free-fall time is short and turbulent and magnetic stresses are weak. The timescale associated with magnetic braking, on the other hand, is initially considerably larger than the free-fall time, though drops considerably between $10-30$~pc. Particularly on scales of $0.1-1$~pc, it is close to the Reynolds timescale and contributes significantly to the redistribution of angular momentum. The timescale of inward advection of the angular momentum is typically found to be somewhat smaller, though fluctuating, compared to the Reynolds and magnetic braking timescales, thereby explaining that the angular momentum still increases in the central region of the collapse. Overall our results thus show that Reynolds and magnetic stresses considerably contribute to the redistribution of angular momentum and reduce the total amount of angular momentum that would be present in pure hydrodynamical runs, though are not sufficient to fully compensate for the advection of angular momentum provided by infall.

\section{Discussion and conclusions} \label{sec:conc}

Using the cosmological high-resolution (magneto-)hydrodynamical simulations of gravitational collapse in atomic cooling halos from L23, we have analyzed here the evolution of angular momentum considering the Reynolds and magnetic torques as well as the inward transport of the angular momentum via convection. Both in hydrodynamical and magneto-hydrodynamical runs, the angular momentum on scales below $1$~pc is found to increase significantly over a timescale of $\sim1$~Myr, but more strongly so in the purely hydrodynamical runs. The magnetic field strength increases significantly over time and while the magnetic torques are initially insignificant, they provide a relevant contribution to the Reynolds torques after about $0.5$~Myrs. Both terms together however do not fully compensate for the inward transport of angular momentum via convection, so that the angular momentum in the center nonetheless keeps increasing.

The dynamical role of the magnetic field with respect to fragmentation has been noted already by L23, as considerably more fragments have formed in the pure hydrodynamical runs, even though many of them are subsequently merging. Here, we show via the comparison of the magnetic and the thermal Jeans mass that indeed on scales of $0.1-10$~pc, the magnetic fields are sufficiently strong to considerably suppress fragmentation. This is similar to results found e.g. by \citet{Sharda20} in the context of smaller halos. Together with the additional contribution to the angular momentum transport, these results explain how the magnetic field aids to reduce fragmentation and favors the formation of a single central object. Such objects are expected to evolve towards a supermassive protostar \citep[e.g.][]{Hosokawa2012,Hosokawa2013,Schleicher13, Haemmerle2019}, which will subsequently contract into supermassive black holes via the General Relativistic instability.

\cite{Pid70} proposed that the gravitational collapse and differential rotation may account for the observed galactic fields within the Hubble timescale. \cite{Rat95} based on timescale arguments anticipated that magnetic braking may remove angular momentum    during the formation of the first cosmic structures. \cite{Pand19} employed an analytical model to study the impact of magnetic braking on the angular momentum in halos of $\rm 10^7-10^9~M_{\odot}$. They found that comoving large scale magnetic fields of strength $\geq$ 0.1 nG are needed to remove the angular momentum from gas clouds. They assumed constant densities of 1, 10 and 100 times the background density and therefore their results do not remain valid on scales below the viral radius of the halo. Also, their toy model does not capture the 3D dynamical effects such as mergers, shocks and turbulent motions which further amplify magnetic fields. Our findings show that even weak seed fields of the order of $\rm 10^{-19}$ G (comoving, at z=150) can be efficiently amplified by the small scale dynamo and in the presence of accretion shocks and significantly contribute to braking the rotation of gas clouds at later times along with Reynold torques.   Therefore, the results of \cite{Pand19} should be considered as an upper limit on the strength of B field required for magnetic braking.

The case of rotating supermassive stars was investigated by \citet{Uchida2017}, finding that angular momentum leads to the formation of a torus surrounding the rotating black hole. In case of the additional presence of magnetic fields, collapse will further lead to the launching of jets  consistent with the typical duration of long gamma-ray bursts \citep{Butler2018, Sun2018}. The presence of magnetic fields during the formation of the first black holes may thus give rise to direct observational implications, and relates early black hole formation to already known observed phenomena.

Our choice of the initial B field is about two order of magnitude smaller than the lower limit inferred from Blazer observations at galactic scales. If we were to employ a stronger initial B field, it would further strengthen our main findings by suppressing fragmentation and efficiently transporting angular momentum by exerting magnetic torques. All in all, it will support the formation of DCBHs. To investigate magnetic braking during the formation of DCBHs, we studied here pristine halos of a few times $\rm10^7~M_{\odot}$ which are considered as embryos of DCBHs. As the rotational velocity scales with halo mass, relatively weaker and stronger B fields will be required to induce magnetic braking in both smaller and larger halos  ($\rm \geq 10^8~M_{\odot}$), respectively,  as found by \cite{Pand19}. However, this needs to be investigated in detail in future works. The strength of magnetic field in our simulations at kpc scales is smaller than observed galactic fields as simulations are only evolved for 1.5 Myr. Studies exploring magnetic fields in Milky Way like galaxies show that e-folding time of about 100 Myr is required until saturation occurs with typical galactic field strength of a 10-50 $\rm \mu$G \citep{Pak17}. Previous works also show that initial magnetic field strength is irrelevant due to the rapid amplification by the small scale dynamo and shocks \citep{Pak14,Latif14mag,Mari16}. Therefore,  if we were to evolve our simulations for a few hundred Myrs they will reproduce observed galactic fields.

\section*{Acknowledgements}

MAL thanks the UAEU for funding via UPAR grants No. 31S390 and 12S111. DRGS gratefully acknowledges support by the ANID BASAL projects ACE210002 and FB21003,  the Millenium Nucleus NCN19-058 (TITANs) as well as via Fondecyt Regular (project code 1201280). DRGS thanks for funding via the  Alexander von Humboldt - Foundation, Bonn, Germany.

\bibliography{smbhs}

\begin{thebibliography}{}
\expandafter\ifx\csname natexlab\endcsname\relax\def\natexlab#1{#1}\fi
\providecommand{\url}[1]{\href{#1}{#1}}
\providecommand{\dodoi}[1]{doi:~\href{http://doi.org/#1}{\nolinkurl{#1}}}
\providecommand{\doeprint}[1]{\href{http://ascl.net/#1}{\nolinkurl{http://ascl.net/#1}}}
\providecommand{\doarXiv}[1]{\href{https://arxiv.org/abs/#1}{\nolinkurl{https://arxiv.org/abs/#1}}}

\bibitem[{{Baym} {et~al.}(1996){Baym}, {B{\"o}deker}, \& {McLerran}}]{Baym96}
{Baym}, G., {B{\"o}deker}, D., \& {McLerran}, L. 1996, \prd, 53, 662,
  \dodoi{10.1103/PhysRevD.53.662}

\bibitem[{{Beck}(2007)}]{Beck07}
{Beck}, R. 2007, \aap, 470, 539, \dodoi{10.1051/0004-6361:20066988}

\bibitem[{{Beck} {et~al.}(1999){Beck}, {Ehle}, {Shoutenkov}, {Shukurov}, \&
  {Sokoloff}}]{Beck99}
{Beck}, R., {Ehle}, M., {Shoutenkov}, V., {Shukurov}, A., \& {Sokoloff}, D.
  1999, \nat, 397, 324, \dodoi{10.1038/16861}

\bibitem[{{Brandenburg} \& {Subramanian}(2005)}]{Brandenburg2005}
{Brandenburg}, A., \& {Subramanian}, K. 2005, \physrep, 417, 1,
  \dodoi{10.1016/j.physrep.2005.06.005}

\bibitem[{{Butler} {et~al.}(2018){Butler}, {Lima}, {Baumgarte}, \&
  {Shapiro}}]{Butler2018}
{Butler}, S.~P., {Lima}, A.~R., {Baumgarte}, T.~W., \& {Shapiro}, S.~L. 2018,
  \mnras, 477, 3694, \dodoi{10.1093/mnras/sty834}

\bibitem[{{de Souza} \& {Opher}(2010)}]{Souza2010}
{de Souza}, R.~S., \& {Opher}, R. 2010, \prd, 81, 067301,
  \dodoi{10.1103/PhysRevD.81.067301}

\bibitem[{{Federrath} {et~al.}(2011){Federrath}, {Sur}, {Schleicher},
  {Banerjee}, \& {Klessen}}]{Fed11}
{Federrath}, C., {Sur}, S., {Schleicher}, D.~R.~G., {Banerjee}, R., \&
  {Klessen}, R.~S. 2011, \apj, 731, 62, \dodoi{10.1088/0004-637X/731/1/62}

\bibitem[{{Galli} {et~al.}(2006){Galli}, {Lizano}, {Shu}, \&
  {Allen}}]{Galli2006}
{Galli}, D., {Lizano}, S., {Shu}, F.~H., \& {Allen}, A. 2006, \apj, 647, 374,
  \dodoi{10.1086/505257}

\bibitem[{{Grasso} \& {Rubinstein}(2001)}]{Gras01}
{Grasso}, D., \& {Rubinstein}, H.~R. 2001, \physrep, 348, 163,
  \dodoi{10.1016/S0370-1573(00)00110-1}

\bibitem[{{Grete} {et~al.}(2019){Grete}, {Latif}, {Schleicher}, \&
  {Schmidt}}]{Grete19}
{Grete}, P., {Latif}, M.~A., {Schleicher}, D.~R.~G., \& {Schmidt}, W. 2019,
  \mnras, 487, 4525, \dodoi{10.1093/mnras/stz1568}

\bibitem[{{Haemmerl{\'e}} {et~al.}(2019){Haemmerl{\'e}}, {Meynet}, {Mayer},
  {Klessen}, {Woods}, \& {Heger}}]{Haemmerle2019}
{Haemmerl{\'e}}, L., {Meynet}, G., {Mayer}, L., {et~al.} 2019, \aap, 632, L2,
  \dodoi{10.1051/0004-6361/201936716}

\bibitem[{{Hennebelle} {et~al.}(2016){Hennebelle}, {Commer{\c{c}}on},
  {Chabrier}, \& {Marchand}}]{Hennebelle2016}
{Hennebelle}, P., {Commer{\c{c}}on}, B., {Chabrier}, G., \& {Marchand}, P.
  2016, \apjl, 830, L8, \dodoi{10.3847/2041-8205/830/1/L8}

\bibitem[{{Hirano} \& {Machida}(2022)}]{Hir22}
{Hirano}, S., \& {Machida}, M.~N. 2022, \apjl, 935, L16,
  \dodoi{10.3847/2041-8213/ac85e0}

\bibitem[{{Hirano} {et~al.}(2021){Hirano}, {Machida}, \& {Basu}}]{Hir21}
{Hirano}, S., {Machida}, M.~N., \& {Basu}, S. 2021, \apj, 917, 34,
  \dodoi{10.3847/1538-4357/ac0913}

\bibitem[{{Hosokawa} {et~al.}(2013){Hosokawa}, {Yorke}, {Inayoshi}, {Omukai},
  \& {Yoshida}}]{Hosokawa2013}
{Hosokawa}, T., {Yorke}, H.~W., {Inayoshi}, K., {Omukai}, K., \& {Yoshida}, N.
  2013, \apj, 778, 178, \dodoi{10.1088/0004-637X/778/2/178}

\bibitem[{{Hosokawa} {et~al.}(2012){Hosokawa}, {Yoshida}, {Omukai}, \&
  {Yorke}}]{Hosokawa2012}
{Hosokawa}, T., {Yoshida}, N., {Omukai}, K., \& {Yorke}, H.~W. 2012, \apjl,
  760, L37, \dodoi{10.1088/2041-8205/760/2/L37}

\bibitem[{{Janka}(2002)}]{Janka2002}
{Janka}, H.-T. 2002, in Lighthouses of the Universe: The Most Luminous
  Celestial Objects and Their Use for Cosmology, ed. M.~{Gilfanov},
  R.~{Sunyeav}, \& E.~{Churazov}, 357, \dodoi{10.1007/10856495_56}

\bibitem[{{Kahniashvili} {et~al.}(2010){Kahniashvili}, {Tevzadze}, {Sethi},
  {Pandey}, \& {Ratra}}]{Kahn10}
{Kahniashvili}, T., {Tevzadze}, A.~G., {Sethi}, S.~K., {Pandey}, K., \&
  {Ratra}, B. 2010, \prd, 82, 083005, \dodoi{10.1103/PhysRevD.82.083005}

\bibitem[{{Kulsrud}(1971)}]{Kulsrud1971}
{Kulsrud}, R.~M. 1971, \apj, 163, 567, \dodoi{10.1086/150801}

\bibitem[{{Latif} \& {Schleicher}(2016)}]{ls15}
{Latif}, M.~A., \& {Schleicher}, D.~R.~G. 2016, \aap, 585, A151,
  \dodoi{10.1051/0004-6361/201527266}

\bibitem[{{Latif} {et~al.}(2016){Latif}, {Schleicher}, \&
  {Hartwig}}]{Latif2016}
{Latif}, M.~A., {Schleicher}, D.~R.~G., \& {Hartwig}, T. 2016, \mnras, 458,
  233, \dodoi{10.1093/mnras/stw297}

\bibitem[{{Latif} {et~al.}(2023){Latif}, {Schleicher}, \& {Khochfar}}]{Latif23}
{Latif}, M.~A., {Schleicher}, D. R.~G., \& {Khochfar}, S. 2023, \apj, 945, 137,
  \dodoi{10.3847/1538-4357/acbcc2}

\bibitem[{{Latif} {et~al.}(2014){Latif}, {Schleicher}, \&
  {Schmidt}}]{Latif14mag}
{Latif}, M.~A., {Schleicher}, D.~R.~G., \& {Schmidt}, W. 2014, \mnras,
  \dodoi{10.1093/mnras/stu357}

\bibitem[{{Latif} {et~al.}(2013{\natexlab{a}}){Latif}, {Schleicher}, {Schmidt},
  \& {Niemeyer}}]{Latif13m}
{Latif}, M.~A., {Schleicher}, D.~R.~G., {Schmidt}, W., \& {Niemeyer}, J.
  2013{\natexlab{a}}, \mnras, 432, 668, \dodoi{10.1093/mnras/stt503}

\bibitem[{{Latif} {et~al.}(2013{\natexlab{b}}){Latif}, {Schleicher}, {Schmidt},
  \& {Niemeyer}}]{Latif13c}
---. 2013{\natexlab{b}}, \mnras, 433, 1607, \dodoi{10.1093/mnras/stt834}

\bibitem[{{Marinacci} \& {Vogelsberger}(2016)}]{Mari16}
{Marinacci}, F., \& {Vogelsberger}, M. 2016, \mnras, 456, L69,
  \dodoi{10.1093/mnrasl/slv176}

\bibitem[{{Neronov} \& {Vovk}(2010)}]{Nero10}
{Neronov}, A., \& {Vovk}, I. 2010, Science, 328, 73,
  \dodoi{10.1126/science.1184192}

\bibitem[{{Pakmor} {et~al.}(2014){Pakmor}, {Marinacci}, \& {Springel}}]{Pak14}
{Pakmor}, R., {Marinacci}, F., \& {Springel}, V. 2014, \apjl, 783, L20,
  \dodoi{10.1088/2041-8205/783/1/L20}

\bibitem[{{Pakmor} {et~al.}(2017){Pakmor}, {G{\'o}mez}, {Grand}, {Marinacci},
  {Simpson}, {Springel}, {Campbell}, {Frenk}, {Guillet}, {Pfrommer}, \&
  {White}}]{Pak17}
{Pakmor}, R., {G{\'o}mez}, F.~A., {Grand}, R. J.~J., {et~al.} 2017, \mnras,
  469, 3185, \dodoi{10.1093/mnras/stx1074}

\bibitem[{{Pandey} {et~al.}(2019){Pandey}, {Sethi}, \& {Ratra}}]{Pand19}
{Pandey}, K.~L., {Sethi}, S.~K., \& {Ratra}, B. 2019, \mnras, 486, 1629,
  \dodoi{10.1093/mnras/stz939}

\bibitem[{{Piddington}(1970)}]{Pid70}
{Piddington}, J.~H. 1970, Australian Journal of Physics, 23, 731,
  \dodoi{10.1071/PH700731}

\bibitem[{{Planck Collaboration} {et~al.}(2016){Planck Collaboration}, {Ade},
  {Aghanim}, {Arnaud}, {Ashdown}, {Aumont}, {Baccigalupi}, {Banday},
  {Barreiro}, {Bartlett}, \& et~al.}]{P16}
{Planck Collaboration}, {Ade}, P.~A.~R., {Aghanim}, N., {et~al.} 2016, \aap,
  594, A13, \dodoi{10.1051/0004-6361/201525830}

\bibitem[{{Pudritz} {et~al.}(2012){Pudritz}, {Hardcastle}, \&
  {Gabuzda}}]{Pud12}
{Pudritz}, R.~E., {Hardcastle}, M.~J., \& {Gabuzda}, D.~C. 2012, \ssr, 169, 27,
  \dodoi{10.1007/s11214-012-9895-z}

\bibitem[{{Pudritz} \& {Silk}(1989)}]{Pudritz1989}
{Pudritz}, R.~E., \& {Silk}, J. 1989, \apj, 342, 650, \dodoi{10.1086/167625}

\bibitem[{{Ratra} {et~al.}(1995){Ratra}, {Kulsrud}, \& {Quillen}}]{Rat95}
{Ratra}, B., {Kulsrud}, R.~M., \& {Quillen}, A.~C. 1995, Princeton preprint
  PUPT-1535 and POP-620

\bibitem[{{Schleicher} {et~al.}(2009){Schleicher}, {Galli}, {Glover},
  {Banerjee}, {Palla}, {Schneider}, \& {Klessen}}]{Schleicher2009}
{Schleicher}, D. R.~G., {Galli}, D., {Glover}, S. C.~O., {et~al.} 2009, \apj,
  703, 1096, \dodoi{10.1088/0004-637X/703/1/1096}

\bibitem[{{Schleicher} {et~al.}(2013){Schleicher}, {Palla}, {Ferrara}, {Galli},
  \& {Latif}}]{Schleicher13}
{Schleicher}, D.~R.~G., {Palla}, F., {Ferrara}, A., {Galli}, D., \& {Latif}, M.
  2013, \aap, 558, A59, \dodoi{10.1051/0004-6361/201321949}

\bibitem[{{Schleicher} {et~al.}(2010){Schleicher}, {Spaans}, \&
  {Glover}}]{Schleicher10}
{Schleicher}, D.~R.~G., {Spaans}, M., \& {Glover}, S.~C.~O. 2010, \apjl, 712,
  L69, \dodoi{10.1088/2041-8205/712/1/L69}

\bibitem[{{Schober} {et~al.}(2012){Schober}, {Schleicher}, {Federrath},
  {Glover}, {Klessen}, \& {Banerjee}}]{Schobera}
{Schober}, J., {Schleicher}, D., {Federrath}, C., {et~al.} 2012, \apj, 754, 99,
  \dodoi{10.1088/0004-637X/754/2/99}

\bibitem[{{Sethi} \& {Subramanian}(2005)}]{Sethi2005}
{Sethi}, S.~K., \& {Subramanian}, K. 2005, \mnras, 356, 778,
  \dodoi{10.1111/j.1365-2966.2004.08520.x}

\bibitem[{{Sharda} {et~al.}(2020){Sharda}, {Federrath}, \&
  {Krumholz}}]{Sharda20}
{Sharda}, P., {Federrath}, C., \& {Krumholz}, M.~R. 2020, \mnras, 497, 336,
  \dodoi{10.1093/mnras/staa1926}

\bibitem[{{Sheikhnezami} \& {Fendt}(2022)}]{Sheikhnezami2022}
{Sheikhnezami}, S., \& {Fendt}, C. 2022, \apj, 925, 161,
  \dodoi{10.3847/1538-4357/ac3f31}

\bibitem[{{Shu} {et~al.}(2007){Shu}, {Galli}, {Lizano}, {Glassgold}, \&
  {Diamond}}]{Shu2007}
{Shu}, F.~H., {Galli}, D., {Lizano}, S., {Glassgold}, A.~E., \& {Diamond},
  P.~H. 2007, \apj, 665, 535, \dodoi{10.1086/519678}

\bibitem[{{Sigl} {et~al.}(1997){Sigl}, {Olinto}, \& {Jedamzik}}]{Sigl97}
{Sigl}, G., {Olinto}, A.~V., \& {Jedamzik}, K. 1997, \prd, 55, 4582,
  \dodoi{10.1103/PhysRevD.55.4582}

\bibitem[{{Silk} \& {Langer}(2006)}]{Silk2006}
{Silk}, J., \& {Langer}, M. 2006, \mnras, 371, 444,
  \dodoi{10.1111/j.1365-2966.2006.10689.x}

\bibitem[{{Sun} {et~al.}(2018){Sun}, {Ruiz}, \& {Shapiro}}]{Sun2018}
{Sun}, L., {Ruiz}, M., \& {Shapiro}, S.~L. 2018, \prd, 98, 103008,
  \dodoi{10.1103/PhysRevD.98.103008}

\bibitem[{{Sur} {et~al.}(2010){Sur}, {Schleicher}, {Banerjee}, {Federrath}, \&
  {Klessen}}]{Sur10}
{Sur}, S., {Schleicher}, D.~R.~G., {Banerjee}, R., {Federrath}, C., \&
  {Klessen}, R.~S. 2010, \apjl, 721, L134, \dodoi{10.1088/2041-8205/721/2/L134}

\bibitem[{{Turk} {et~al.}(2012){Turk}, {Oishi}, {Abel}, \& {Bryan}}]{Turk2012}
{Turk}, M.~J., {Oishi}, J.~S., {Abel}, T., \& {Bryan}, G.~L. 2012, \apj, 745,
  154, \dodoi{10.1088/0004-637X/745/2/154}

\bibitem[{{Uchida} {et~al.}(2017){Uchida}, {Shibata}, {Yoshida}, {Sekiguchi},
  \& {Umeda}}]{Uchida2017}
{Uchida}, H., {Shibata}, M., {Yoshida}, T., {Sekiguchi}, Y., \& {Umeda}, H.
  2017, \prd, 96, 083016, \dodoi{10.1103/PhysRevD.96.083016}

\bibitem[{{Widrow} {et~al.}(2012){Widrow}, {Ryu}, {Schleicher}, {Subramanian},
  {Tsagas}, \& {Treumann}}]{Widrow2012}
{Widrow}, L.~M., {Ryu}, D., {Schleicher}, D. R.~G., {et~al.} 2012, \ssr, 166,
  37, \dodoi{10.1007/s11214-011-9833-5}

\end{thebibliography}
\bibliographystyle{aasjournal}

\end{document}